\documentclass{SciPost}

\hypersetup{
    colorlinks,
    linkcolor={red!50!black},
    citecolor={blue!50!black},
    urlcolor={blue!80!black},
    pdftitle={Dancing in the dark: probing Dark Matter through the dynamics of eccentric binary pulsars}, 
}

\usepackage[bitstream-charter]{mathdesign}
\usepackage{graphicx}
\usepackage{txfonts}
\usepackage{hyperref}
\usepackage{comment}
\usepackage{siunitx}
\usepackage{amsmath}

\DeclareMathOperator\erf{erf}
\DeclareMathOperator{\Cin}{Cin}

\usepackage{tikz}
\usetikzlibrary{shapes}

\usepackage{color}

\fancypagestyle{SPstyle}{
\fancyhf{}
\lhead{\colorbox{scipostblue}{\bf \color{white} ~SciPost Physics }}
\rhead{{\bf \color{scipostdeepblue} ~Submission }}

\fancyfoot[C]{\textbf{\thepage}}
}

\begin{document}

\pagestyle{SPstyle}

\begin{center}{\Large \textbf{\color{scipostdeepblue}{
Dancing in the dark: probing Dark Matter through the dynamics of eccentric binary pulsars
}}}\end{center}

\begin{center}\textbf{
Giorgio Nicolini\textsuperscript{1,2,3$\star$},
Andrea Maselli\textsuperscript{4,5$\dagger$} and
Miguel Zilhão\textsuperscript{2,3,6$\ddagger$}
}\end{center}

\begin{center}
{\bf 1} Montana State University, Bozeman, MT 59717, USA \\
{\bf 2} Department of Mathematics, University of Aveiro, 3810-193 Aveiro, Portugal\\
{\bf 3} Center for Research and Development in Mathematics and Applications (CIDMA), \\
        University of Aveiro, 3810-193 Aveiro, Portugal\\
{\bf 4} Gran Sasso Science Institute (GSSI), I-67100 L’Aquila, Italy\\
{\bf 5} INFN, Laboratori Nazionali del Gran Sasso, I-67100 Assergi, Italy\\
{\bf 6} Department of Physics, University of Aveiro, 3810-193 Aveiro, Portugal
\\[\baselineskip]
$\star$ \href{mailto:email1}{\small giorgionicolini@montana.edu}\,,\quad
$\dagger$ \href{mailto:andrea.maselli@gssi.it}{\small andrea.maselli@gssi.it}\,,\quad
$\ddagger$ \href{mailto:mzilhao@ua.pt}{\small mzilhao@ua.pt}
\end{center}

\section*{\color{scipostdeepblue}{Abstract}}
\textbf{\boldmath{%
We investigate the dynamics of eccentric binary 
pulsars embedded in dark matter environments. While 
previous studies have primarily focused on circular 
orbits in collisionless dark matter halos, we extend 
this framework to eccentric systems and explore their 
interaction with ultralight scalar fields. Adopting 
a perturbative approach, we compute the modifications 
to the orbital period induced by dark matter-driven 
dynamical friction. Our results show that orbital 
eccentricity amplifies the imprints of non-vacuum 
environments on binary dynamics, underscoring the 
potential of such systems as sensitive probes for 
dark matter signatures.
}}

\vspace{\baselineskip}

\vspace{10pt}
\noindent\rule{\textwidth}{1pt}
\tableofcontents
\noindent\rule{\textwidth}{1pt}
\vspace{10pt}


%
\section{Introduction}\label{sec:intro}
%
Astrophysical systems do not evolve in isolation. Rather, 
they interact with complex environments, evolving together 
with a multitude of matter and fields, possibly of unknown 
nature. Such environments influence the dynamics of 
compact objects through mechanisms like accretion, 
gravitational pull and drag, as well as tidal and gas 
torques, leaving distinctive imprints across the 
electromagnetic and gravitational wave (GW) spectra~\cite{Gondolo:1999ef,Barausse:2014tra}. 
Such imprints are potentially observable by next-generations 
GW detectors across a wide range of frequencies~\cite{Abac:2025saz,Corsi:2024vvr,LISA:2024hlh,Luo:2025ewp,Ajith:2024mie}, 
and by radio telescopes used for pulsar timing 
arrays~\cite{Kramer:2004hd}.
Accurately characterizing these imprints is crucial --- not only 
to avoid systematic biases caused by the incorrect modeling of 
vacuum templates but also to unlock a deeper understanding of 
the environments in which black holes and neutron stars evolve. 
These imprints encode key properties of the surrounding medium, 
such as its density, viscosity, and spatial scales.

Probing the nature of dark matter (DM) is among the most 
ambitious scientific goals that can be pursued through the 
study of these astrophysical settings and their interplay 
with compact objects. DM remains one of the most enigmatic 
components of the universe, constituting approximately 27\% 
of its total mass-energy density~\cite{Planck:2018vyg}, 
while continuing to elude direct detection. 
A major challenge in understanding DM lies in the extraordinary uncertainty surrounding its possible mass scales, which span 
nearly 80 orders of magnitude. Proposed candidates range from 
ultralight bosons, such as axions or fuzzy DM with masses as 
low as $10^{-22}~\unit{eV}$\cite{Marsh:2015xka}, to weakly 
interacting massive particles (WIMPs) at the GeV to TeV scale~\cite{Bertone:2018krk,Roszkowski:2017nbc}, and even 
primordial black holes, whose masses span from sub-planetary 
to supermassive scales~\cite{Carr:2020xqk}. 

In dense regions near galactic centers, DM tends to cluster around black holes, forming dense ``spikes'' with density peaks near the black hole horizon~\cite{Gondolo:1999ef, Sadeghian:2013laa}. These overdensities modify the dynamics of nearby binaries by altering the gravitational potential and inducing drag forces.
Estimates of the local DM density in the Solar neighborhood are typically $0.3 \pm 0.1~\unit{GeV.cm^{-3}}$~\cite{Read:2014qva}, but localized density enhancements, such as subhalos or ultracompact clumps, could amplify DM interactions in certain environments. 

Previous studies of DM effects on compact binaries can be broadly divided into those involving purely gravitational interactions and those invoking additional non-gravitational couplings. Works focusing on the gravitational influence of ambient DM --- through dynamical friction, density spikes, or surrounding halos --- which affect the orbital dynamics without requiring any direct coupling to the binary components, include~\cite{Pani:2015qhr,Caputo:2017zqh,AbhishekChowdhuri:2023rfv}. Scenarios based on ultralight scalar fields may also introduce non-gravitational effects, such as resonances or time-dependent scalar interactions, which can produce observable signatures only in the presence of direct couplings~\cite{Blas:2019hxz,Kus:2024vpa}.
Alternative DM models, including fermionic particles and massive scalar or vector fields, have also been analyzed for their effects on binary pulsar dynamics~\cite{Gomez:2019mtl,Seymour:2020yle}, and the role of spin-2 fields and oscillating ultralight vector fields in these models was investigated in Refs.~\cite{Armaleo:2019gil,LopezNacir:2018epg}.
Additionally, accreting pulsar-black hole binaries have been suggested as probes for DM properties, with studies providing insights into more exotic DM models~\cite{Akil:2023kym,Goldman:2020nee}.

Herein, we will focus exclusively on gravitational drag effects, without assuming any direct coupling between DM and the binary constituents.
In fact, observations of double pulsars have been proposed as a novel avenue for investigating the properties of DM in the pulsar's vicinity~\cite{Pani:2015qhr}. DM particles influence the binary dynamics through dynamical friction, which drags the components and induces a secular change in the orbital period. In~\cite{Pani:2015qhr} this effect was investigated for collisionless DM particles, focusing on binaries with circular orbits. 
There have also been numerous works on the effects of dynamical friction and its interaction with black holes~\cite{Vicente:2022ivh,Traykova:2021dua,Traykova:2023qyv, Kocsis:2011dr, Speeney:2024mas,Cardoso:2022nzc,Zhong:2023xll}.

In this work, we continue the investigation of DM detection 
via binary pulsar timing, extending the analysis of~\cite{Pani:2015qhr} in two key directions. First, we incorporate sources on 
eccentric orbits, as observed in double pulsar systems~\cite{Fonseca:2014qla,MataSanchez:2020pys} and 
white dwarf–pulsar binaries~\cite{Antoniadis:2010eq}. 
Orbital eccentricity is expected to amplify the relative 
velocity near periastron, thereby enhancing dynamical 
friction effects~\cite{Bar-Or:2012zrb}. Second, we consider 
the influence of ultralight DM, which can substantially modify the orbital decay rate compared to the 
collisionless case. 
For each scenario, we calculate the induced changes in the 
orbital period as functions of the binary parameters and 
evaluate their observability with current and upcoming 
instruments.

\section{Dynamical evolution of binary pulsars}\label{sec:formalism}

We analyze the interaction between the motion of a 
binary pulsar and its DM environment using a perturbative 
approach based on the method of osculating orbits, 
which is particularly well-suited to study the 
evolution of binary systems under arbitrary 
perturbing forces~\cite{poisson2014gravity}, including 
those induced by DM overdensities. At zeroth order, 
the binary system in vacuum follows Keplerian dynamics, while environmental effects are treated as first-order perturbations. In this section, we provide a brief 
overview of the formalism, extending the results 
of \cite{Pani:2015qhr} to generic orbits and arbitrary 
perturbing forces.

\subsection{Kepler problem}

In the absence of perturbations, 
the motion of a binary system with total 
mass $M = m_1 + m_2$ can be described using two reference frames: (a) the 
{\it fundamental} (or Galactic) frame, a fixed frame with Cartesian coordinates 
$\{X, Y, Z\}$; and (b) the {\it co-rotating} frame, which rotates with a virtual 
particle of mass $\mu = \frac{m_1 m_2}{M}$  around the system's center 
of mass. The latter is defined by the coordinates $\{x, y, z\}$, such that the fixed orbital plane coincides with the $xy$ plane, and  is 
accompanied by the constant basis vectors $\boldsymbol{e}_x, \boldsymbol{e}_y,  \boldsymbol{e}_z$ and by the 
time-dependent basis $\boldsymbol{n}, \boldsymbol{\lambda}, \boldsymbol{e}_z$.

The description of the motion relative to the fundamental frame is given in terms 
of six orbital elements:  (i) the semi-latus rectum $p = h^2/(GM)$, where 
$h$ is the magnitude of the reduced angular momentum, defined as 
$\boldsymbol{h} = \boldsymbol{r} \times \boldsymbol{v}$;  
(ii) the eccentricity $e$, which for bound orbits satisfies $0 \leq e < 1$;  
(iii) the orbital inclination $\iota$, which specifies the angle between the 
$z$-direction of the orbital plane and the $Z$-direction of the fixed fundamental frame;
(iv) the longitude of the ascending node $\Omega$, defined as the angle 
between the $X$-axis and the line of nodes, i.e., the intersection of the orbital 
plane with the fundamental frame;  (v) the longitude of pericenter $\omega$, 
which determines the orientation of the orbit within the orbital plane and is 
defined as the angle between the pericenter and the line of nodes, measured in 
the orbital plane; and (vi) the true anomaly $f = \phi - \omega$, that represents the angle between the separation vector and the direction 
of the pericenter, with $\phi$ being the azimuthal angle 
of the body in the orbital plane as measured from the ascending node. In this parametrization, the semi-major axis of the binary orbit is given 
by $a = \frac{p}{1 - e^2}$.

In the co-rotating frame, the displacement vector between 
the two component masses $\boldsymbol{r} \coloneqq \boldsymbol{r}_2 - \boldsymbol{r}_1$ and its time derivative 
$\boldsymbol{v}$, read
\begin{equation}
\label{eq:r_and_v_def}
    \boldsymbol{r} = r \, \boldsymbol{n}\ , \qquad 
    \boldsymbol{v} = \dot r \, \boldsymbol{n} + v_\perp \, \boldsymbol{\lambda}\ ,
\end{equation}
where
\begin{equation}
\label{eq:kepler_orbit}
    r = \frac{p}{1 + e \cos f}\ , \qquad
    \dot r = \sqrt{ \frac{GM}{p}} e \sin f\ , \qquad
    v_\perp =  \sqrt{ \frac{GM}{p}} \left(1 + e \cos f \right)\ , 
\end{equation}
describe the usual Keplerian orbit.
The time-dependent basis $\boldsymbol{n}, \boldsymbol{\lambda}, \boldsymbol{e}_z$ can also be expressed in the fundamental frame as
\begin{subequations}
\begin{align}
    \boldsymbol{n}  = &\left[ \cos \Omega \cos(\omega + f) - \cos \iota \sin \Omega \sin(\omega +f) \right] \boldsymbol{e}_X+\left[ \sin \Omega \cos(\omega + f)\right.
    \nonumber\\
    &\left. + \cos \iota \cos \Omega \sin(\omega +f) \right] \boldsymbol{e}_Y+ \sin \iota \sin(\omega + f) \, \boldsymbol{e}_Z\ ,\label{eq:corot-frame-kepler1} \\
    \boldsymbol{\lambda} = &\left[ -\cos \Omega \sin(\omega + f) - \cos \iota \sin \Omega \cos(\omega +f) \right] \boldsymbol{e}_X+\left[-\sin \Omega \sin(\omega + f)\right. \nonumber\\
    &\left. + \cos \iota \cos \Omega \cos(\omega +f) \right] \boldsymbol{e}_Y+ \sin \iota \cos(\omega + f) \, \boldsymbol{e}_Z\ , \label{eq:corot-frame-kepler2}\\
    \boldsymbol{e}_z =& \sin \iota \sin \Omega \, \boldsymbol{e}_X
    - \sin \iota \cos \Omega \, \boldsymbol{e}_Y
    + \cos \iota \, \boldsymbol{e}_Z\ .\label{eq:corot-frame-kepler3}
\end{align}
\end{subequations}
%

\subsection{Perturbed Kepler problem}

Additional forces perturbing the zeroth-order Keplerian motion 
cause the orbital elements to evolve over time. These changes can 
be described with the method of osculating orbits, in which the 
perturbed trajectory is treated as a sequence of instantaneous 
Keplerian orbits that match the true motion at every instant. 
Assuming that the relative acceleration between the two bodies 
is generically given by
\begin{equation}
   \boldsymbol{a} \coloneqq \boldsymbol{a}_2 - \boldsymbol{a}_1 = - \frac{GM}{r^2} \boldsymbol{n} + \boldsymbol{F}\ ,\label{eq:acc}
\end{equation}
we decompose the perturbing force per unit mass $\boldsymbol{F}$ in the co-rotating frame 
as
\begin{equation}\label{eq:forceproj}
    \boldsymbol{F} = \mathcal{R} \, \boldsymbol{n} + \mathcal{S} \, \boldsymbol{\lambda} + \mathcal{W} \, \boldsymbol{e}_z \ ,
\end{equation}
with
\begin{equation}
\mathcal{R}=\boldsymbol{F}\cdot\boldsymbol{n} \,, \quad 
\mathcal{S}=\boldsymbol{F}\cdot\boldsymbol{\lambda} \,,\quad
\mathcal{W}=\boldsymbol{F}\cdot\boldsymbol{e}_z \, .
\end{equation}
Following~\cite{poisson2014gravity}, at first order in perturbation theory, the 
evolution of the orbital elements is then given by
\begin{subequations}
\label{eq:osculating}
\begin{align}
    \frac{dp}{df} & \simeq 2 \frac{p^3}{G M} \frac{1}{(1+e \cos f)^3} \mathcal{S}(f), \label{eq:dpdf} \\
    \frac{de}{df} & \simeq \frac{p^2}{GM} \left[ 
       \frac{\sin f}{(1+e \cos f)^2} \mathcal{R}(f) 
       + \frac{ 2 \cos f + e(1+e \cos^2 f) }{(1+e \cos f)^3} \mathcal{S}(f)
    \right]\ , \label{eq:dedf}  \\
    \frac{d \iota}{df} & \simeq \frac{p^2}{GM} 
        \frac{ \cos(\omega + f) }{(1+e \cos f)^3} \mathcal{W}(f),  \\
    \sin \iota \frac{d \Omega}{df} & \simeq \frac{p^2}{GM} 
        \frac{ \sin(\omega + f) }{(1+e \cos f)^3} \mathcal{W}(f),   \\
    \frac{d \omega}{df} & \simeq \frac{1}{e} \frac{p^2}{GM} \left[ 
      - \frac{\cos f}{(1+e \cos f)^2} \mathcal{R}(f)
      + \frac{2 + e \cos f}{(1+e \cos f)^3} \sin f \, \mathcal{S}(f)
      - e \cot \iota \frac{\sin(\omega + f)}{(1+e \cos f)^3} \mathcal{W}(f)
    \right]\ ,
\end{align}
\end{subequations}
with
\begin{equation}
    \frac{dt}{df} \simeq \sqrt{\frac{p^3}{GM}} \frac{1}{(1+e \cos f)^2}
    \left[ 
        1 - \frac{1}{e} \frac{p^2}{GM} \left( 
                \frac{\cos f}{(1+e \cos f)^2} \mathcal{R}(f)
                - \frac{2 + e \cos f}{(1+e \cos f)^3} \sin f \, \mathcal{S}(f)
        \right)
    \right]\ ,
\end{equation}
where the right-hand side of these equations is understood to be evaluated at the (zeroth-order) Keplerian solution.
The coefficients $\mathcal{R}$, $\mathcal{S}$, and $\mathcal{W}$ depend on the specific form of the 
perturbing force. In the next section we will 
compute these coefficients for different dynamical friction models.

Equations~\eqref{eq:dpdf} and~\eqref{eq:dedf} allow us to obtain the $f$ derivative of the orbital period.
With these equations, Kepler's third law $P_b^2 = \dfrac{4\pi^2}{GM}\,a^3$ and 
the relation between $p$ and the semi-major axis $a$, 
$p = a\,(1 - e^2)$, we can write
\begin{equation}
    \label{eq:dPdf}
    \frac{dP_b}{df} \simeq 3 P_b \frac{p^2}{GM} \frac{1}{ (1-e^2) (1 + e \cos f)^2}
    \left[
        \frac{1+e \cos f \ (2 + e^2 \cos f)}{1+e \cos f} \mathcal{S}(f)
        + e \sin f \ \mathcal{R}(f)
    \right]
\ .
\end{equation}
We are interested in \emph{secular changes} of a generic 
orbital parameter $\zeta^a$ and its time derivative, obtained 
by averaging them over an orbit:
\begin{equation}
\Delta \zeta^a = \int_0^{2\pi} \frac{d \zeta^a}{df} df \,, \quad 
\langle \dot \zeta^a \rangle \coloneqq \frac{\Delta \zeta^a}{P_b}\ .
\end{equation}
For the orbital period change \eqref{eq:dPdf} this yields
\begin{equation}
    \label{eq:Pdot-sec}
    \langle \dot P_b \rangle = 3 \frac{p^2}{GM (1-e^2)} 
    \int_0^{2\pi}
    \left[
        \frac{1+e \cos f \ (2 + e^2 \cos f)}{(1+e \cos f)^3} \mathcal{S}(f)
        + \frac{e \sin f}{(1 + e \cos f)^2} \ \mathcal{R}(f)
    \right]
    \, df\ .
\end{equation}
The time derivative of the orbital period, $\dot{P}_b$, 
is an observable quantity measurable through pulsar 
timing, encoding information about the astrophysical 
environment in which pulsars evolve. In the following, 
we use Eq.~\eqref{eq:Pdot-sec} as a figure of merit 
to quantify the impact of DM on their orbital dynamics.

\section{The perturbing forces}

\subsection{Dynamical friction for collisionless dark matter}
\label{sec:chandraDM}

The dynamical friction force $\boldsymbol{F}_i^\text{DF}$ experienced by 
an object of mass $m_i$ in linear motion through a homogeneous 
collisionless medium of density $\rho_\textrm{DM}$ is given by~\cite{Chandrasekhar:1943ys,RevModPhys.21.383} 
\begin{equation} \label{eq:dyn-fric}
    \boldsymbol{F}_i^\text{DF} = -4\pi\rho_\textrm{DM}\lambda\frac{G^2\,m_i^2}{\tilde{v}_i^3} \left[\erf(X_i) - \frac{2 X_i}{\sqrt{\pi}}e^{-X_i^2} \right]\boldsymbol{\tilde{v}}_i\ ,
\end{equation}
where $\boldsymbol{\tilde{v}}_i = \boldsymbol{\dot{r}}_i + \boldsymbol{v}_w$ denotes the object's velocity relative to the wind of DM particles, with $\boldsymbol{r}_i$ representing the object's coordinate vector and $\boldsymbol{v}_w$ the rotational velocity of the binary around the galaxy (which, neglecting the rotational velocity of the DM halo, is opposite to the velocity of the wind of DM particles relative
to the center of mass), 
$X_i \coloneqq \dfrac{\tilde{v}_i}{\sqrt{2}\sigma}$, 
with $\sigma$ identifying the dispersion of the 
DM Maxwellian velocity distribution, and 
$\lambda \approx 20_{-10}^{+10}$ is the Coulomb logarithm\footnote{We fix $\lambda=20$ throughout this work for consistency with the existing literature (see e.g.~\cite{Pani:2015qhr}). In any case, since this value appears as an overall factor, all our results can be appropriately rescaled.}.
Now, we consider a binary system of masses $m_1\geq m_2$ 
and total mass $M$, affected by dynamical friction in a 
non-vacuum environment. The displacement 
vector between the two bodies is $\boldsymbol{r} = \boldsymbol{r}_2 - \boldsymbol{r}_1$.
Following the approach of~\cite{Pani:2015qhr}, we
recast Eq.~\eqref{eq:dyn-fric}
\begin{equation}
\boldsymbol{F}_{i}^{\rm DF}= -A b_i \frac{m_i^2}{M} \boldsymbol{\tilde{v}}_i \, ,\quad \ {\rm where} \quad 
A = 4\pi\rho_\textrm{DM}\lambda G^2 M \, ,  \quad 
b_i = \dfrac{1}{\tilde{v}_i^3}\left[ \erf(X_i) - \dfrac{2X_i}{\sqrt{\pi}}e^{-X_i^2}\right]\ , 
\label{eq:FDF_chandra}
\end{equation}
and the equations of motion for $\boldsymbol{r}$ and the center-of-mass position 
$\boldsymbol{R}$
are given by
\begin{align}
\dot{\boldsymbol{v}} &=  - \frac{G M}{r^3}\,\boldsymbol{r} + a_1\eta\,\boldsymbol{v} + a_2(\boldsymbol{v}_w + \boldsymbol{V})\ ,\label{eqm-v}\\ 
\dot{\boldsymbol{V}} &= a_2\eta\,\boldsymbol{v} + a_3(\boldsymbol{v}_w + \boldsymbol{V})\ ,\label{eqm-V}            
\end{align}
where $\eta=\mu/M=m_1 m_2/M^2$ is the symmetric mass ratio, and 
the coefficients $a_1$, $a_2$, and $a_3$ are defined as  
\begin{align}
    a_1 = -A(b_1 + b_2)\ ,  \qquad
    a_2 = \frac{A}{2} \left(b_1\Delta_{+} + b_2\Delta_{-}\right)\ , \qquad
    a_3 = -\frac{A}{4} \left(b_1\Delta_{+}^2 + b_2\Delta_{-}^2\right) \ ,
\end{align}
where $\Delta_\pm = \Delta \pm 1$, $\Delta = \sqrt{1 - 4\eta}$. 
Equations~\eqref{eqm-v}-\eqref{eqm-V} show that, 
in the presence of a non-vacuum environment, the binary's center of mass 
experiences an acceleration due to dynamical friction.  

To apply the osculating orbit method, we must separate the zeroth-order 
(unperturbed) and first-order (perturbed) components of the binary's 
dynamics. To this aim we introduce the bookkeeping parameter 
$\epsilon\coloneqq \frac{\rho_\text{DM}}{M} L^3\sim\rho_\text{DM} G P_b^2$, 
which determines the ratio between friction and the gravitational 
force, where $L$ is the characteristic orbital separation of the binary, and we use it as perturbative parameter for the method of osculating orbits. For $\epsilon\rightarrow0$ we recover the 
vacuum binary configuration.
We then expand all vectors in Eqs.~\eqref{eqm-v}-\eqref{eqm-V} as  $\boldsymbol{u} = \boldsymbol{u}^{(0)} + \epsilon \, \boldsymbol{u}^{(1)}$.
At zeroth order, $\boldsymbol{V}^{(0)} = 0$; keeping only first-order terms in $\epsilon$, Eq.~\eqref{eqm-v} takes the form
\[
\dot{\boldsymbol{v}} =  - \frac{G M}{r^2}\,\boldsymbol{n} + a_1\eta\,\boldsymbol{v}^{(0)} + a_2\boldsymbol{v}_w \,,
\]
which does not depend on the movement of the center-of-mass $\boldsymbol{V}$. 
Comparing this with Eq.~\eqref{eq:acc}, we can straightforwardly identify the perturbing force per unit mass, $\boldsymbol{F}$, as  
\begin{equation} 
    \label{eq:pert-force}
    \boldsymbol{F} = a_1\eta\,\boldsymbol{v}^{(0)} + a_2\boldsymbol{v}_w \,.
\end{equation}

To decompose the perturbing force~\eqref{eq:pert-force} in the 
form of Eq.~\eqref{eq:forceproj}, we first use the zeroth-order 
Keplerian solution to orient the Galactic frame such that $\Omega^{(0)} = \omega^{(0)} = \iota^{(0)} = 0$. This choice simplifies the projection of vectors and is made without loss of generality, as physical observables are invariant under coordinate rotation. Then, using Eqs.~\eqref{eq:corot-frame-kepler1}-\eqref{eq:corot-frame-kepler3}, 
we obtain the basis vectors $(\boldsymbol{n}, \boldsymbol{\lambda}, \boldsymbol{e}_z)$:
\begin{equation}
\boldsymbol{n}^{(0)} = (\cos f^{(0)},~\sin f^{(0)},~0), \qquad
\boldsymbol{\lambda}^{(0)} = (-\sin f^{(0)},~\cos f^{(0)},~0), \qquad
\boldsymbol{e}^{(0)}_z = (0,~0,~1)\ .
\end{equation}
Given that the velocity vector is expressed as
\begin{equation}
\boldsymbol{v}^{(0)} = \dot{r}^{(0)} \boldsymbol{n}^{(0)} + v_{\perp}^{(0)}\boldsymbol{\lambda}^{(0)}\ ,
\end{equation}
we obtain
\begin{equation}
\boldsymbol{v}^{(0)} = \left( \dot{r}^{(0)} \cos f^{(0)}
		+v_{\perp}^{(0)}\sin f^{(0)} \right) \boldsymbol{e}_X +
\left(\dot{r}^{(0)} \sin f^{(0)} + v{\perp}^{(0)} \cos f^{(0)} \right) \boldsymbol{e}_Y\ .
\end{equation}
Moreover, we decompose the velocity vector $\boldsymbol{v}_w$ as
\begin{equation}
\boldsymbol{v}_w = v_w \, (\sin\beta\cos\alpha,~\sin\beta\sin\alpha,~\cos\beta)\ ,
\end{equation}
which we assume to be constant over the observation period. 
Here, $\alpha$ and $\beta$ are the azimuthal and polar angles defining the direction of $\boldsymbol{v}_w$ in the Galactic 
reference frame. The projection of the perturbing force is 
then given by
\begin{subequations}
\begin{align}
\mathcal{R} &= \boldsymbol{F} \cdot \boldsymbol{n}^{(0)} = a_1 \eta \, {\dot r}^{(0)}
+ a_2 v_w \sin \beta \cos(f - \alpha)\ , \label{eq:R} \\
\mathcal{S} &= \boldsymbol{F} \cdot \boldsymbol{\lambda}^{(0)} = a_1 \eta \, v^{(0)}_\perp
- a_2 v_w \sin \beta \sin(f-\alpha)\ , \label{eq:S} \\
\mathcal{W} &= \boldsymbol{F} \cdot \boldsymbol{e}_z^{(0)} = a_2 v_w \cos \beta\ . \label{eq:W}
\end{align}
\end{subequations}
Using the (Keplerian) expressions for the radial and orthogonal components of the velocity
\begin{equation}
\label{eq:vperp_rdot}
v_{\perp}^{(0)} = \sqrt{\frac{G M}{p}} \, (1 + e\cos f)\ , \qquad
\dot r^{(0)} = \sqrt{\frac{GM}{p}} e \sin f\ ,
\end{equation}
we finally obtain the explicit form of $\mathcal{S}(f)$ and $\mathcal{R}(f)$, 
required to compute the shift in the orbital period,
\begin{subequations}
\label{eq:SR_2}
\begin{align}
{\cal S}(f) & = a_1 \eta\sqrt{\frac{GM}{p}}(1+e\cos f) - a_2 v_w\sin\beta\sin(f-\alpha) \ ,\label{eq:S_2} \\
\mathcal{R} (f) & = a_1 \eta\sqrt{\frac{GM}{p}} e \sin f + a_2 v_w\sin\beta\cos(f-\alpha) \ .
\label{eq:R_2}
\end{align}
\end{subequations}

\subsubsection{Validity of the linear motion approximation}
\label{sec:validity_chandra}

Equation~\eqref{eq:dyn-fric} was derived for linear motion, but it can also be applied to a binary provided that one can neglect the interaction with the companion's wake. Following~\cite{1990ApJ...359..427B} and \cite{Pani:2015qhr}, this expression should hold provided that the characteristic size of the wake~$L$ is much smaller than the semi-minor axis of the binary orbit~$b$. Estimating $L$ from the size of the gravitational sphere of influence as in~\cite{1990ApJ...359..427B} and \cite{Pani:2015qhr}, we have the condition $m_\textrm{DM} \sigma^2 \sim G m_i m_\textrm{DM} / L $. Using Kepler's law, the condition $L \ll b$ can then be written as
\begin{equation}
    P_b \gg \frac{G m_i}{\sigma^3} \frac{1}{(1-e^2)^{\frac{3}{4}}} 
    \sim \frac{0.46}{(1-e^2)^{\frac{3}{4}}} \left( \frac{m_i}{M_\odot} \right) 
    \left( \frac{150~\unit{km/s}}{\sigma} \right)^3~\unit{day} .
    \label{eq:validity_chandra}
\end{equation}
We therefore see that, even for very high values of eccentricity of $e \simeq 0.98$, we have
\[
    P_b \gg 5.2 \left( \frac{m_i}{M_\odot} \right) 
    \left( \frac{150~\unit{km/s}}{\sigma} \right)^3~\unit{day} ,
\]
which validates the use of Eq.~\eqref{eq:dyn-fric} for orbital periods of $P_b \gtrsim 100$~days. 
This threshold is satisfied by various known binary pulsars (see, e.g., the ATNF Pulsar Catalogue~\cite{Manchester:2004bp}).

\subsection{Dynamical friction for ultra-light dark matter}
\label{sec:sfDM}

Considering now an ultralight scalar-field DM model, the friction force was first derived in~\cite{Hui:2016ltb} for non-relativistic velocities, extended in~\cite{Vicente:2022ivh} to the relativistic case and also extracted and explored numerically in~\cite{Traykova:2021dua, Traykova:2023qyv}. 
This friction arises from the gravitational response of 
the scalar field medium --- a coherent wave --- as the object 
moves through it, generating a trailing density distortion 
(or wake) that acts back on the object.
The expression takes the form 
\begin{equation}
\label{eq:FDFSF0}
\boldsymbol{F}_i^\text{SF} = -4\pi\rho_\textrm{DM}\frac{G^2 m_i^2}{\tilde v_i^3} {\cal C}_i \, \boldsymbol{\tilde v}_i
\end{equation}
where $\mathcal{C}_i$ can be written in a closed formed fashion in different regimes of validity.

Defining the scalar field mass parameter $\mu_\text{SF} \coloneqq \frac{m_\text{SF} c}{\hbar}$ (where $1/\mu_\text{SF}$ is its reduced Compton wavelength), we introduce the parameter 
$\alpha_s \coloneqq \mu_\text{SF}  \frac{G M}{c^2}$, which 
encodes the effective gravitational coupling between the DM field and the binary.
In the limit $\alpha_s \ll 1$, $ \alpha_s \frac{c}{\tilde{v}_i} \ll 1$, $\mathcal{C}_i$ takes the form~\cite{Hui:2016ltb,Traykova:2021dua}
\begin{equation}
\label{Ci-ULDM}
\mathcal{C}_i \simeq \Cin\left(\mu_\text{SF} \frac{\tilde{v}_i}{c} b_\text{max} \right) 
    + \frac{\sin\left(\mu_\text{SF} \frac{\tilde{v}_i}{c} b_\text{max} \right)}{\mu_\text{SF} \frac{\tilde{v}_i}{c} b_\text{max}} - 1
\end{equation}
where
$\Cin(z) = \int_0^z (1 - \cos t)/t~dt$ is the cosine integral 
and $b_\text{max} = 2R$, given by twice the cutoff 
radius, represents an infrared cutoff 
which regularizes the long-range nature of the interaction and depends on the effective size of the perturbed DM region.
In the following, we set the cutoff radius $R$ to the semi-minor axis 
of the binary orbit, which represents the characteristic spatial scale 
over which each component of the binary perturbs the surrounding DM 
environment~\cite{Hui:2016ltb,Traykova:2021dua,Traykova:2023qyv}.
Similarly to previous sections, we can rewrite Eq.~\eqref{eq:FDFSF0} as 
\begin{equation}
\label{eq:DFSF}
\boldsymbol{F}_i^\text{SF}=-A^\text{SF} \frac{m_i^2}{M} b^\text{SF}_i \boldsymbol{\tilde v}_i, 
\quad A^\text{SF} = 4\pi\rho_\textrm{DM} G^2 M, \quad 
b^\text{SF}_i = \dfrac{{\cal C}_i}{\tilde v_i^3} \ .
\end{equation}  

We can follow once again the same steps as before to 
obtain the projections of the perturbing force $\mathcal{S}(f)$ and $\mathcal{R}(f)$ due to scalar-field DM by taking Eqs.~\eqref{eq:SR_2} and replacing the coefficients $a_{1,2}$ with their corresponding values from Eq.~\eqref{eq:DFSF}.

\section{Changes in the orbital period due to Dark Matter}
\label{sec:constraining}

We now apply the expressions derived in the previous section 
to estimate the change in the orbital period of a binary 
system under various DM models. We remark that,
while the dynamical friction formulas were originally 
developed for linear motion, as discussed in Sec.~\ref{sec:validity_chandra}, 
they provide reasonable approximations for systems with 
sufficiently large orbital periods. We therefore adopt 
them here as proxies to model systems in bound orbital configurations satisfying condition~\eqref{eq:validity_chandra}.\footnote{The bound derived in Sec.~\ref{sec:validity_chandra} is strictly valid for the collisionless DM model. For simplicity, however, we will apply this criterion to the other environmental models we consider.}

The DM density appears as an overall scaling factor in the expressions 
for the perturbing forces~\eqref{eq:FDF_chandra} and \eqref{eq:DFSF}, and 
consequently in the expression for the period derivative \eqref{eq:Pdot-sec}. 
For convenience, we will perform all our computations using a 
reference value of $\rho_0 = 1~\unit{GeV.cm^{-3}}$, which is the order of magnitude estimated for the local DM density from recent observations~\cite{deSalas:2020hbh}. 
For the velocity $v_w$ and velocity dispersion $\sigma$ of the DM wind we will explore different values around the typically used order of magnitude of $v_w,\,\sigma \sim 100~\unit{km/s}$~\cite{Cerdeno:2010jj,Lancaster:2019mde,Battaglia:2005rj}.

\begin{figure}[tphb]
\centering
\includegraphics[width=0.85\linewidth]{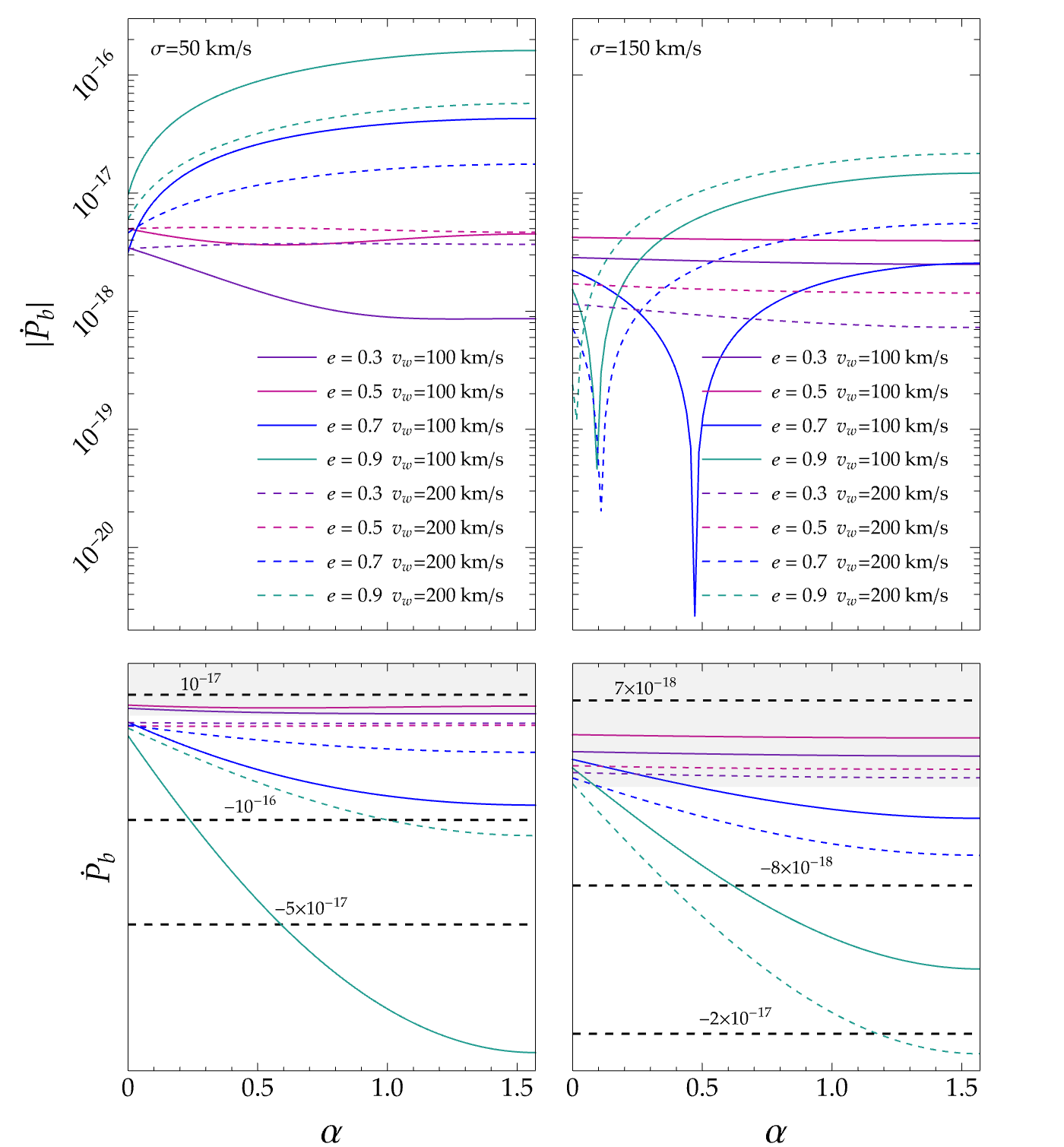} 
\caption{(Top Row) Absolute value of the 
secular change in the orbital period as a 
function of $\alpha$ for binary pulsars with 
$m_1=1.3M_\odot$ and $m_2=0.3M_\odot$ and different 
values of eccentricity. For all configurations 
we assume $\beta=\pi/2$ and $P_b=100$ days. Solid 
and dashed curves refer to DM wind speed of $v_w=100$ 
km/s and $v_w=200$ km/s, while left and right panel 
to velocity dispersion of $\sigma=50$ km/s and 
$\sigma=150$ km/s, respectively. (Bottom Row) 
Secular change in the orbital period for the same 
configurations shown in the top panels. Horizontal 
dashed black lines identify specific values of $\dot{P}_b$. 
Gray and white regions correspond to parameter space in 
which $\dot{P}_b>0$ and $\dot{P}_b<0$, respectively.
}
\label{Fig1}
\end{figure}

We begin our analysis by examining the collisionless DM scenario described in Sec.~\ref{sec:chandraDM}. Figure~\ref{Fig1} illustrates the variation in the orbital period as a function of the angle $\alpha$ for binary systems with $P_b = 100$ days on eccentric orbits. Here we fix $\beta = \pi/2$, corresponding to a DM wind flowing parallel to the orbital plane. We explore different values for both the DM wind velocity $v_w$ and the velocity dispersion $\sigma$. To facilitate comparison with previous studies, we adopt the same component masses used in \cite{Pani:2015qhr}, namely $m_1 = 1.3 M_\odot$ and $m_2 = 0.3 M_\odot$.\footnote{The results for zero eccentricity match those shown in Fig.~3 of Ref.~\cite{Pani:2015qhr} when rescaled using the DM density $\rho_\textrm{DM} = 2\times 10^3~\unit{GeV.cm^{-3}}$ employed therein.}

In contrast to circular orbits, whose secular variations do 
not depend on $\alpha$~\cite{Pani:2015qhr}, eccentric binaries 
exhibit a more structured dependence of $\dot{P}_b$. For $\sigma = 50\,\unit{km/s}$ (left panel of Fig.~\ref{Fig1}), systems with relatively large eccentricity, $e \gtrsim 0.5$, display an 
orbital-period derivative whose magnitude increases monotonically as 
$\alpha$ varies over the interval $[0,\pi/2]$. Increasing 
the velocity dispersion (right panel) introduces additional 
structure: in this case, $|\dot{P}_b|$ tends to decrease 
initially and then rise again as $\alpha$ increases, with 
the location of the turning point depending on $e$.  
For smaller eccentricities, this behavior is absent irrespective 
of $\sigma$, and $\dot{P}_b$ remains approximately constant 
or shows a mild decrease as $\alpha$ varies. Larger values 
of $\sigma$ also lead to an overall suppression of $\dot{P}_b$. 
By comparison, increasing $v_w$ produces only a modest 
reduction in $|\dot{P}_b|$, without modifying the qualitative 
trends discussed above.

Overall, in contrast to the circular case, the dependence of $\dot{P}_b$ on the system parameters for elliptical orbits underscores the importance of accounting for orbital 
eccentricity when assessing the impact of dark matter 
on the dynamics of binary pulsars.

\begin{figure}[thpb]
    \centering
       \includegraphics[width=1\linewidth]{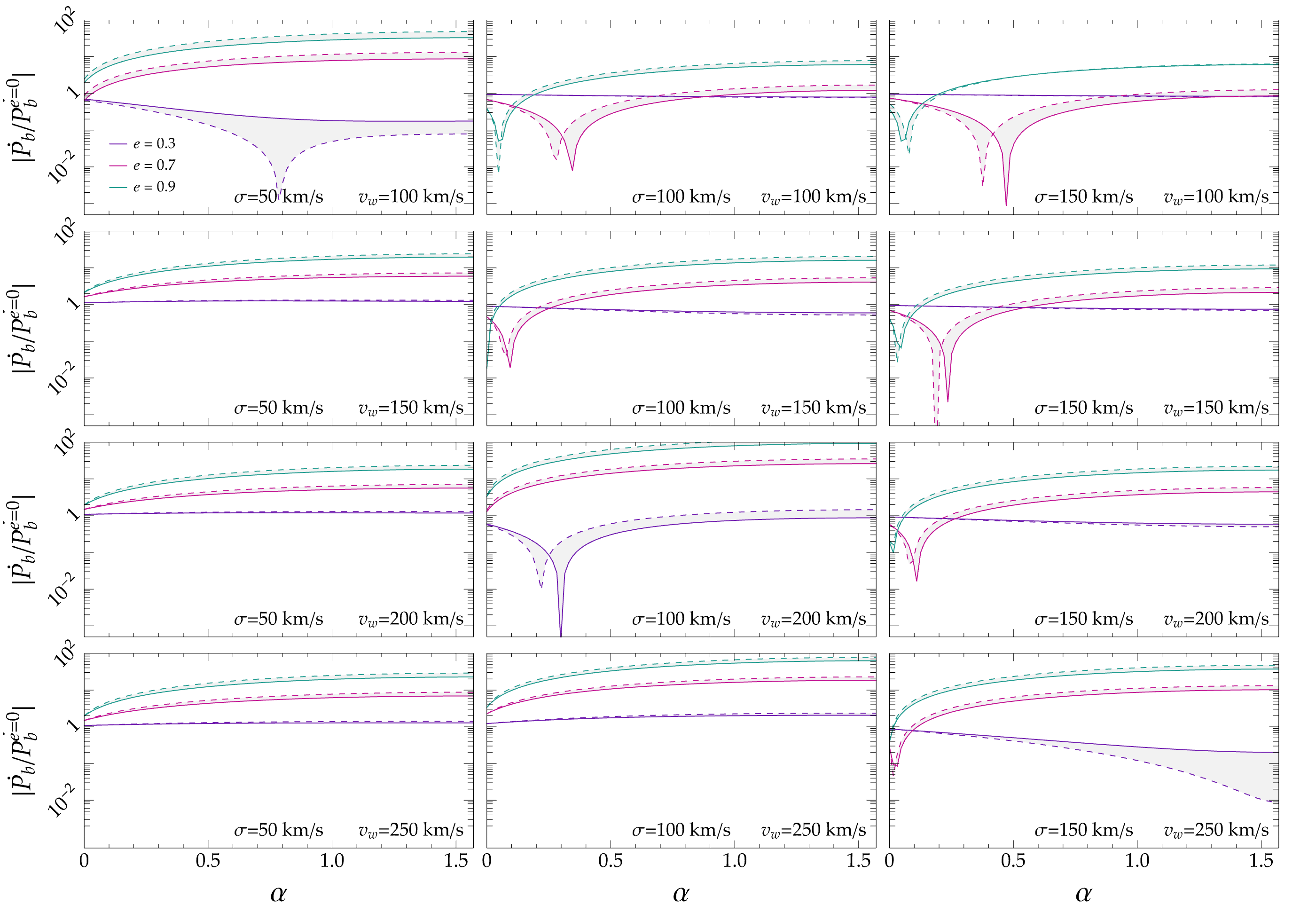} 
\caption{Secular change in the orbital period as given 
from Eq.~\eqref{eq:Pdot-sec} for binaries on eccentric 
orbits, normalized to the circular case, as a function 
of the angle $\alpha$. Each panel corresponds to a different 
choice of the  DM wind velocity $v_w$ and of the velocity 
dispersion $\sigma$. Values of $\dot{P}_{b}$ are  obtained 
for a binary pulsar with component masses $m_1=1.3M_\odot$ 
and $m_2=0.3M_\odot$. The lower solid (upper dashed) curve of each 
colored band identify binaries with orbital period of $P_b=100$ 
($P_b=200$) days. For all panels we fix $\beta=\pi/2$.}
\label{Fig2}
\end{figure}

Figure~\ref{Fig2} further compares the secular evolution of $P_b$ for circular and eccentric binaries across different orbital and DM parameters as computed from Eq.~\eqref{eq:Pdot-sec}. 
In Fig.~\ref{Fig2}, 
solid and dashed curves, enclosed within shaded bands, correspond to orbital periods of $P_b = 100$ days and $P_b = 200$ days, respectively.
For a fixed value of $\alpha$, the orbital-period derivative 
in eccentric systems can differ from the circular-orbit case by 
up to two orders of magnitude. At the lowest velocity dispersion 
considered, $\sigma = 50~\unit{km/s}$, $\dot{P}_b$ may exceed 
its circular counterpart by roughly one order of magnitude for 
$e = 0.7$, and by two orders or magnitude for $e = 0.9$. For the smallest 
eccentricity examined, $e = 0.3$, the ratio $|\dot{P}_b/\dot{P}_b^{e=0}|$ 
falls below unity, decreasing monotonically with $\alpha$ when 
$P_b = 200$, and developing a turning point when $P_b = 100$.

Consistent with Fig.~\ref{Fig1}, variations in $v_w$ have only a 
limited impact, although $|\dot{P}_b/\dot{P}_b^{e=0}|$ remains 
$\gtrsim 1$ for all eccentricities when $v_w$ is sufficiently low. 
Allowing for $\sigma > 50~\unit{km/s}$ at fixed $v_w$ 
introduces additional structure: the ratio now develops a turning point 
and can change sign as $\alpha$ varies, even for $e = 0.7$ and $e = 0.9$. 
This behavior gradually disappears as $v_w$ increases, with variations in the orbital period of eccentric systems consistently exceeding 
those of circular ones.  Configurations with $e = 0.3$ generally 
yield $|\dot{P}_b/\dot{P}_b^{e=0}| \sim 1$, although for the largest 
values of $\sigma$ and $v_w$ considered, the trend reverses and 
changes in $P_b$ become larger in the circular case.

\begin{figure}[thpb]
    \centering
	\includegraphics[width=1\linewidth]{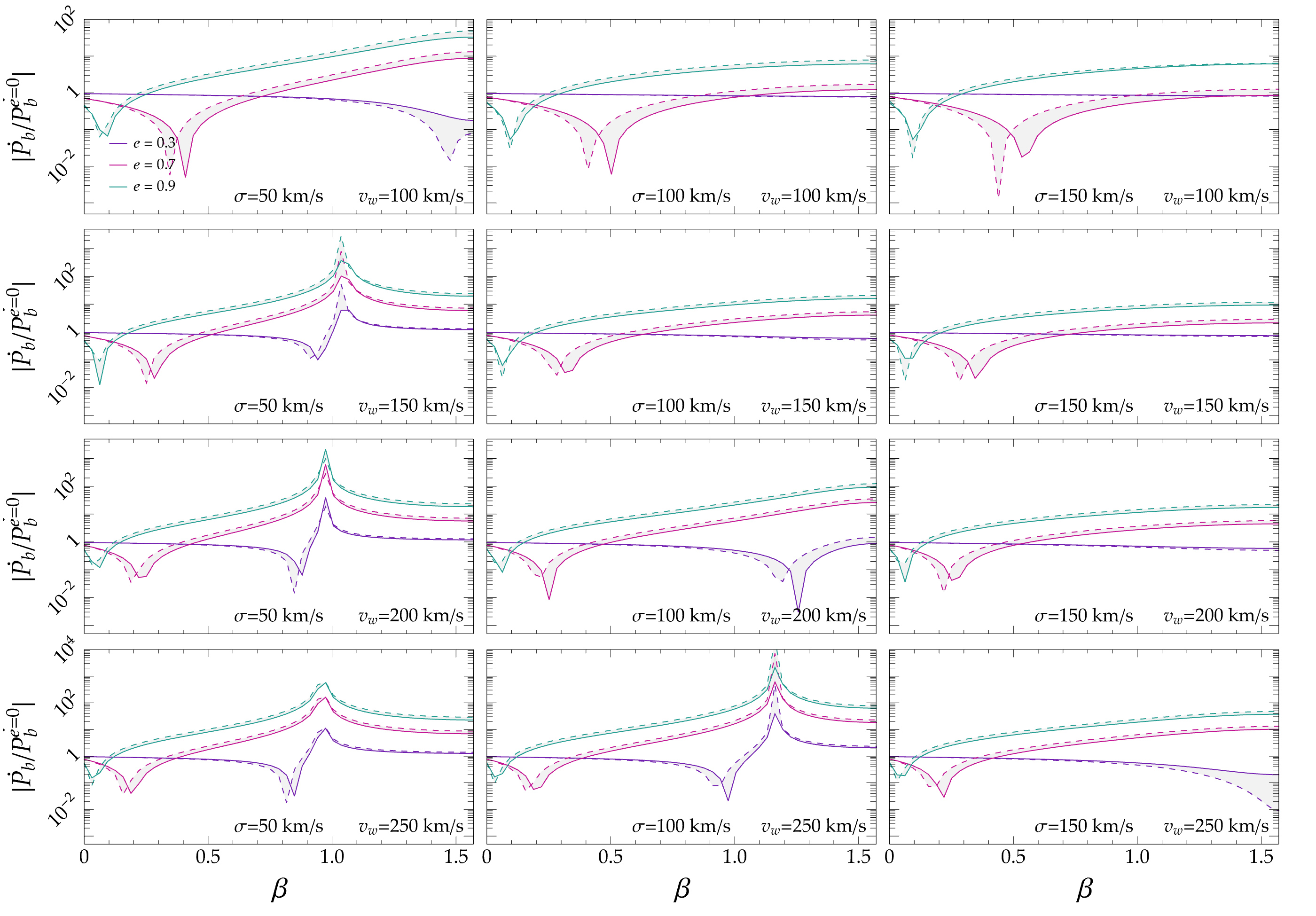} 
    \caption{Same as Fig.~\ref{Fig2} but fixing $\alpha=\pi/2$ and varying $\beta$.}
    \label{Fig3}
\end{figure}

Figure~\ref{Fig3} shows the values of ${\dot P}_b$ as a function of 
$\beta$, fixing $\alpha = \pi/2$, for the same component masses and 
orbital configurations considered in Fig.~\ref{Fig1}. The overall 
behavior of ${\dot P}_b$ is consistent with the trends discussed 
above: the eccentricity produces a clear splitting in the secular 
change for a given $\beta$, leading to substantial deviations from 
the circular case examined in \cite{Pani:2015qhr}. Variations in $\beta$, 
however, generate a more intricate structure than those in $\alpha$, 
with the ratio $|\dot{P}_b/\dot{P}_b^{e=0}|$ developing multiple turning 
points as $\beta$ varies.

\begin{figure}[thpb]
    \centering
	\includegraphics[width=0.9\linewidth]{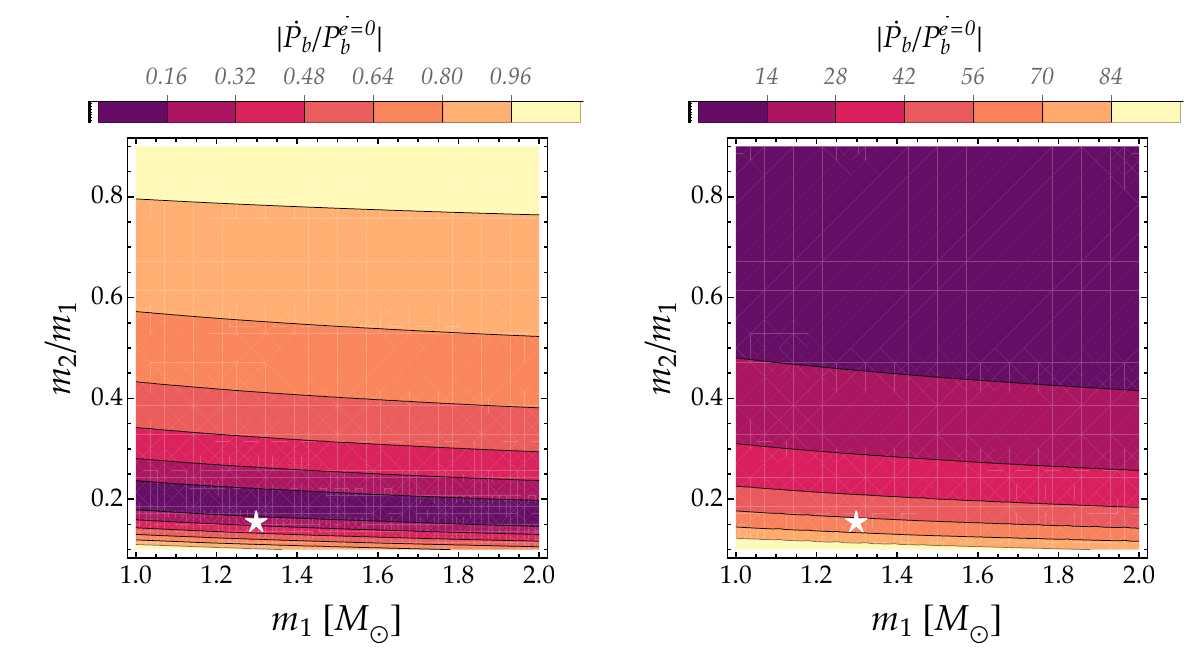} 
    \caption{Density plots for the change in the orbital period for binaries on eccentric orbits (left panel: $e=0.3$; right panel: $e=0.9$), normalized to the circular case, as a function of $m_1$ and the mass ratio $m_2/m_1$. In both panels, we fix $\alpha=\beta=\pi/2$, $P_b = 100$ days, $\sigma = 100~\unit{km/s}$, and $v_w = 250~\unit{km/s}$. The yellow star identifies the case discussed in Figs.~\ref{Fig1}--\ref{Fig2}, with $m_1 = 1.3\,M_\odot$ and $m_2 = 0.3\,M_\odot$.}
    \label{Fig4}
\end{figure}

To assess the dependence of ${\dot P}_b$ on the component masses, 
we varied $m_{1,2}$ and found that their influence on the results 
is generally smaller than an order of magnitude. 
Figure~\ref{Fig4} displays density plots of the ratio $|\dot{P}_b/\dot{P}^{e=0}_b|$ 
as a function of $m_1$ and the mass ratio $q = m_2/m_1$, assuming 
$\alpha = \beta = \pi/2$, $P_b = 100$ days, and $(\sigma, v_w) = (100, 250)\,\unit{km/s}$. The left and right panels correspond to binaries with 
$e = 0.3$ and $e = 0.9$, respectively.

For $e = 0.3$, the ratio $|\dot{P}_b/\dot{P}^{e=0}_b|$ drops well 
below unity as the system becomes increasingly asymmetric, i.e.\ 
for $q \ll 1$. In contrast, for $e = 0.9$, the orbital period changes exceed those of the circular case across the full 
parameter range, with the smallest differences occurring near 
equal masses, $q \sim 1$.

We now turn to the ultra-light dark matter scenario 
described in Sec.~\ref{sec:sfDM}, adopting the 
friction coefficient $\mathcal{C}_i$ given in 
Eq.~\eqref{Ci-ULDM}, which applies in the low-velocity 
and small-$\alpha_s$ regime~\cite{Hui:2016ltb}. In 
this case, the prospects for observing changes in the 
orbital period are less favorable than in the collisionless 
DM scenario. For the component masses considered above, 
$m_1 = 1.3\,M_\odot$ and $m_2 = 0.3\,M_\odot$, and an 
orbital period of $P_b = 100$ days, we find typical 
values $\lvert \dot{P}_b \rvert \lesssim 10^{-20}$ 
for $\mu_{\rm SF} \lesssim 10^{-8}\,\mathrm{m}^{-1}$, 
with the magnitude of the orbital period change increasing 
mildly with the scalar-field mass.

Larger boson masses are limited by the range of validity 
of the small-coupling approximation underlying 
Eq.~\eqref{Ci-ULDM}. The resulting values of 
$\dot{P}_b$ are nearly insensitive to the angles $(\alpha,\beta)$, the velocity $v_w$, and, notably, the binary 
eccentricity. Although somewhat larger effects can be 
obtained for systems with longer orbital periods, 
even for $P_b = 500$ days we find $\lvert \dot{P}_b \rvert \lesssim 10^{-18}$, still with no appreciable dependence 
on the eccentricity.

\section{Final remarks}

In this study, we investigated the dynamics of eccentric 
binary pulsars evolving in DM-rich environments. 
We computed changes in the orbital period due to dynamical 
friction induced by the surrounding medium, modeled either as collisionless or ultralight DM. For each DM 
scenario, we explored the binary parameter space, analyzing how variations in the orbital period depend on both the properties 
of the environment and the orbital configuration. Our results 
show that the impact of DM on the orbital evolution 
of binary pulsars depends sensitively on the underlying microphysical model. While collisionless DM can induce sizable, 
eccentricity-dependent modifications of the orbital period, 
the ultralight scalar-field scenario leads to significantly 
weaker effects, with a reduced sensitivity to orbital 
parameters and system properties.

In the collisionless scenario, orbital eccentricity significantly amplifies the effects of dynamical friction, enhancing them 
by more than an order of magnitude compared to circular orbits. Depending on the relative orientation of the binary with 
respect to the DM wind, $\dot{P}_b$ can change sign, 
with turning points dictated by the eccentricity and by the 
properties of the DM distribution, such as the 
velocity dispersion.

From an observational perspective, it is crucial to account 
for the intrinsic derivative of the orbital period, which includes contributions from kinematic effects due to the relative 
acceleration between the binary pulsar and the Solar System 
barycenter, as well as the energy loss due to gravitational-wave 
emission. The kinematic contributions alone can reach values 
of order $\dot{P}_b^{\rm intr} \sim 10^{-15}$, requiring 
precise knowledge of the pulsar distance for reliable subtraction~\cite{Hu:2025nua}. To isolate the effect of 
dynamical friction, it is therefore advantageous to consider 
systems in which the orbital decay driven by GW emission is subdominant. This typically occurs in binaries with longer orbital periods; 
however, such systems also pose observational challenges, as 
measuring $\dot{P}_b$ with high precision becomes increasingly 
difficult at large orbital periods.

Current pulsar-timing experiments already achieve remarkable 
precision in measuring $\dot{P}_b$ for compact and highly 
relativistic binaries. Indeed, measurements in systems such as 
PSR~B1913+16 and the double pulsar J0737$-$3039A/B report 
fractional uncertainties as low as $\sigma_{\dot{P}_b}/|\dot{P}_b| \sim 10^{-4}$--$10^{-5}$~\cite{Weisberg:2016jye,Kramer:2006nb}. This 
level of precision is not yet attainable for the wider binaries 
considered here, with $P_b \gtrsim 100$ days, for which 
intrinsic orbital-period derivatives are often not reported 
in the ATNF Pulsar Catalogue~\cite{Manchester:2004bp}. 
Nevertheless, this situation is expected to improve 
substantially with next-generation facilities such as the Square 
Kilometre Array (SKA), whose enhanced sensitivity, 
expanded pulsar census, and extended timing baselines are 
expected to enable significantly more precise determinations 
of $\dot{P}_b$ in long-period binaries than currently 
achievable ones~\cite{Kramer:2015dfa,Keane:2015jta}.

Using the binary pulsar PSR~J1302$-$6350~\cite{Manchester:2004bp,Shannon:2013dpa} as a representative example, with parameters $m_1 \simeq 20\,M_\odot$, $m_2 \simeq 1.4\,M_\odot$, eccentricity $e \simeq 0.87$, and orbital period $P_b \simeq 1237~\unit{days}$, our collisionless DM model predicts a variation in the orbital period due to dynamical friction of $\dot P_b \simeq 6.7 \times 10^{-15}$ for $v_w = 250~\unit{km/s}$, and $\dot P_b \simeq 1.8 \times 10^{-14}$ for $v_w = 150~\unit{km/s}$, assuming reference values $\sigma = 50~\unit{km/s}$, $\alpha=\beta=\pi/2$, and $\rho_0 = 1~\unit{GeV.cm^{-3}}$ for the DM density.\footnote{From Eq.~\eqref{eq:FDF_chandra}, $\dot P_b$ scales linearly with the DM density.} These values have the opposite sign and are more than an order of magnitude larger than the orbital decay expected from gravitational-wave emission in this system, $\dot P^{\rm GW}_b \simeq -7.7 \times 10^{-16}$, indicating that dynamical friction could dominate the secular evolution of the orbit in such wide and highly eccentric binaries.

For typical Solar-neighborhood DM densities of 
$\rho_\odot \sim 0.3~\unit{GeV.cm^{-3}}$~\cite{deSalas:2020hbh}, 
the resulting effect on the orbital period would be challenging 
to detect. However, in regions with higher DM densities, 
such as the Galactic Center—where estimates suggest $\rho_\mathrm{DM} \sim 10^2~\unit{GeV.cm^{-3}}$ at distances of 100~pc from the 
center and up to $\rho_\mathrm{DM} \sim 10^3~\unit{GeV.cm^{-3}}$ at 10~pc~\cite{Shen:2023kkm,Sofue:2020rnl}—these effects are 
expected to be amplified and could potentially be measurable~\cite{Caputo:2017zqh}. Additionally, as noted in~\cite{Mishra:2025yaq}, binaries located in regions of 
enhanced DM density may also provide improved sensitivity to accretion-based effects. Recently reported systems such as the millisecond binary pulsar PSR~J1744$-$2946~\cite{Lower:2024sdi}, 
located within 140~pc of the Galactic Center, exemplify 
promising targets for future observational constraints.

Looking ahead, next-generation facilities such as the SKA 
are expected to detect thousands of new pulsars, with the 
goal of uncovering the majority of the Galactic population~\cite{Kramer:2015dfa,Keane:2015jta}. As more 
binary pulsars are discovered, particularly those with 
high eccentricities and/or located in DM-rich 
environments, the prospects for detecting or constraining 
dynamical-friction effects will improve substantially. Our 
results provide a framework for interpreting such future 
measurements and highlight classes of systems in which these 
signals may be within reach of next-generation timing arrays.


\section*{Acknowledgements}
We thank the anonymous 
referee for their valuable comments 
which have improved the quality of 
our manuscript. We also thank Rodrigo~Vicente 
for helpful discussions and 
Paolo~Pani for having carefully read the 
manuscript and for useful comments.
M.Z.\ further thanks Paulo Teixeira and José 
Manuel Mota for discussions.

\paragraph{Funding information}
G.N.\ and M.Z.\ acknowledge financial support by the Center for Research and Development in Mathematics 
and Applications (CIDMA) through Fundaç\~ao para a Ci\^encia e a Tecnologia (FCT) Multi-Annual 
Financing Program for R\&D Units through grants UID/4106/2025 (\url{https://doi.org/10.54499/UID/04106/2025}) and UID/PRR/4106/2025
as well as the following projects: PTDC/FIS-AST/3041/2020 
(\url{http://doi.org/10.54499/PTDC/FIS-AST/3041/2020}); 2022.04560.PTDC 
(\url{https://doi.org/10.54499/2022.04560.PTDC}); and 2022.00721.CEECIND 
(\url{https://doi.org/10.54499/2022.00721.CEECIND/CP1720/CT0001}).
A.M.\ acknowledges financial support from MUR PRIN Grant No.\ 2022-Z9X4XS, funded by the 
European Union --- Next Generation EU.

\bibliography{refs}

\end{document}